\documentclass[review]{elsarticle}

\pdfoutput=1
\UseRawInputEncoding
\usepackage{algorithm}
\usepackage{algorithmic}
\usepackage{multirow}
\usepackage{tabularx}
\usepackage{diagbox}
\usepackage{graphicx}
\usepackage{subfigure}
\usepackage{threeparttable}
\usepackage{amsmath}
\usepackage{amsfonts}
\usepackage{booktabs}
\usepackage{color}
\definecolor{cblue}{RGB}{0,0,160}

\begin{document}
\begin{frontmatter}

\title{A Novel Semi-supervised Meta Learning Method for Subject-transfer Brain-computer Interface}
\author[address1]{Jingcong Li}
\cortext[address1]{This work was supported in part by the National Natural Science Foundation of China (Grant No. 61836003, 61573150). 
}
\author[address1]{Fei Wang\corref{mycorrespondingauthor}}
\cortext[mycorrespondingauthor]{Corresponding author}
\ead{scutauwf@foxmail.com}
\author[address1]{Haiyun Huang}
\author[address2]{Feifei Qi}
\author[address1]{Jiahui Pan}

\address[address1]{School of Software, South China Normal University, Guangzhou, China}
\address[address2]{School of Internet Finance and Information Engineering, Guangdong University of Finance, Guangzhou, China}
\address[address3]{Pazhou Lab, Guangzhou, China}

\begin{abstract}
Brain-computer interface (BCI) provides a direct communication pathway between human brain and external devices. Before a new subject could use BCI, a calibration procedure is usually required. Because the inter- and intra-subject variances are so large that the models trained by the existing subjects perform poorly on new subjects. Therefore, effective subject-transfer and calibration method is essential. In this paper, we propose a semi-supervised meta learning (SSML) method for subject-transfer learning in BCIs. The proposed SSML learns a meta model with the existing subjects first, then fine-tunes the model in a semi-supervised learning manner, i.e. using few labeled and many unlabeled samples of target subject for calibration. It is significant for BCI applications where the labeled data are scarce or expensive while unlabeled data are readily available. To verify the SSML method, three different BCI paradigms are tested: 1) event-related potential detection; 2) emotion recognition; and 3) sleep staging. The SSML achieved significant improvements of over $15\%$ on the first two paradigms and $4.9\%$ on the third. The experimental results demonstrated the effectiveness and potential of the SSML method in BCI applications.
\end{abstract}
\begin{keyword}
Semi-supervised, Meta learning, Transfer learning, Event-related potential, Emotion recognition, Sleep staging.
\end{keyword}

\end{frontmatter}

\section{Introduction}
Brain-computer interface (BCI) is a system to record and decode signals of human brain for the communication with external devices \cite{lotte2015signal}. It provides a direct communication pathway between the paralyzed patients or those with severe motor impairments and the real world. Electroencephalogram (EEG) is widely used in practical BCI systems because EEG is cheap, easy to use, and has relative high temporal resolution among all the other non-invasive techniques. There were many different kinds of EEG-based BCI applications, such as event-related potential detection \cite{hoffmann2008efficient}, emotion recognition \cite{lijp2019}, sleep stage classification \cite{banluesombatkul2021}, etc. For different application scenarios, there are many different kinds of BCI paradigms and different types of EEG signals. For example, affective BCI is to utilize BCI system to detect, process, and respond to the emotional states of subjects \cite{muhl2014survey}. To facilitate the research of EEG-based emotion recognition, SJTU emotion EEG dataset (SEED) was released \cite{duan2013differential}. Many EEG emotion recognition algorithms were evaluated on the SEED dataset. Moreover, EEG-based BCI system could be applied to record, process and monitor the sleep stages of the chronic insomnia \cite{chriskos2021review}. In order to automatically recognize the sleep stages, many machine learning algorithms were proposed and evaluated on a benchmark dataset, i. e. Sleep-EDF dataset \cite{Kemp2000}. BCI platform based on event-related potential (ERP) could be used to investigate age-related changes in brain function \cite{kieffaber2016evaluation}, external associate and communicate devices for disabled subjects \cite{hoffmann2008efficient}, etc. As a matter of fact, advanced EEG decoding techniques are essential for the application and deployment of BCI systems.

In the current BCI applications, the EEG decoding performance is mainly affected by the various EEG signals of different subjects in different time. Specifically, EEG signals rapidly change over times due to the influence of circumstance, the fluctuation of BCI system and physiological/psychological conditions of subjects \cite{lotte2015signal}. In addition, there are large inter-subject and intra-subject variances of EEG signals of different subjects, such as spatial origin, mental conditions, brain shapes, rhythms, feature types, etc \cite{wei2018subject}. As a result, the models trained by the existing subjects or data are unlikely suitable for the other subjects or the same subject in a different time \cite{kwon2020}. Therefore, calibration procedure is required for a new subject before using BCI system.

To calibrate EEG decoding models for new subjects, different kinds of methods were proposed. For example, transfer Learning methods for cross-subject EEG classification in BCI systems had attracted much attention in the past few years \cite{Wu2020}. To effectively deploy EEG-based sensorimotor BCI system, a cross-subject EEG decoding method was proposed \cite{saha2020intra}. Transfer learning method based on riemannian geometry was effective for cross-subject classification in BCI applications \cite{congedo2017}. A new convolutional neural network, termed DynamicNet, achieved state-of-the-art performance of cross-subject classification in motor imagery \cite{zancanaro2021cnn}. Domain adaption technique is able to learn knowledge from source domain and transfer to a new target domain. An easy domain adaptation method was proposed for cross-subject multi-view emotion recognition \cite{CHEN2022}. However, it is difficult and time-consuming to collect sufficient high-quality labeled data for calibration. And the current cross-subject decoding methods could not compensate the large inter- and intra-subject variance of subjects in different circumstances. Therefore, effective subject-transfer and calibrating method is significant.

Meta learning, known as learning to learn is a specific machine learning methodology which aims to learn knowledge or experience from meta data or source domain and effectively learn new tasks in the target domain. Recently, meta learning method was proven to be a powerful tool to capture common knowledge or experiences of different subjects which could be transferred to a new incoming subject for better decoding performance \cite{Ning2021, Li2021, Choi2021}. Usually, a meta learning method will firstly train a meta learner with many labeled samples of source subjects and fine-tune/calibrate meta learner with few-shot labeled samples for a new incoming subject \cite{duan2020ultra, choi2021meta, bhosale2022calibration}. A few-shot relation learning method was proposed for cross-subject motor imagery classification performance \cite{an2020}. The relation learning method was applied to learn representative features of unseen subject categories and how to classify them with limited number of samples. A prototype network based on SPD matrix was proposed for domain adaptation EEG emotion recognition \cite{wang2021}. The proposed prototype network could transfer knowledge by feature adaptation with distribution confusion and sample adaptation with centroid alignment. Siamese neural network is a specific meta learning method based on metric between samples. A multimodal-Siamese neural network was applied for person verification \cite{CHAKLADAR202117}. These studies demonstrated that meta learning methods are promising for subject-transfer EEG decoding and calibration in BCIs.

To reduce the discrepancy between target and source distributions, meta learning method like model agnostic meta-learning (MAML) could be applied for subject-transfer learning \cite{finn_maml2017}. With MAML method, EEG decoding network could be pre-trained by the averaged gradients of source subjects and fine-tune the network with just few-shot labeled samples of a new incoming subject for his/her calibration \cite{banluesombatkul2021, bontonou2021}. However, one major problem of the MAML method is that the calibration procedure using only few samples will cause over-fitting and performance degradation on target subject. Because only a small fraction of labeled samples are unlikely to capture intrinsic data distribution of target subject. Generally, the labeled data are scarce or expensive while unlabeled data are readily available in BCI applications and neuroscience studies. Therefore, semi-supervised learning methods are received much attention.

In this paper, we combine semi-supervised learning and meta learning methods to transfer EEG decoding models for new subjects just with few-shot labeled and many unlabeled samples. The main contributions of this paper can be summarized as follows:
\begin{enumerate}
  \item A novel semi-supervised meta learning (SSML) method is proposed for subject-transfer EEG decoding.
  \item The proposed method is flexible for different BCI paradigms including event-related potential detection, emotion recognition and sleep staging tasks.
  \item The SSML could significantly improve EEG decoding performance on new subjects, e.g. improvements of over $15\%$ in ERP and emotion paradigms and $4.9\%$ in the sleep paradigm.
\end{enumerate}

The remainder of this paper is organized as follows. The proposed SSML method is presented in Section II. In Section III, numerical experiments on event-related potential detection, emotion recognition and sleep staging datasets are carried out. And the performance of the current methods and the proposed method are presented and compared. Some discussions and analysis of the proposed method are presented in Section IV. Conclusions of this paper are given in Section V.

\section{Proposed Method}

As aforementioned, transfer learning cross subjects is significant for BCI applications and neuroscience researches. However, it is quite challenging for subject-transfer EEG decoding due to individual variances of EEG signals, such as mental conditions, brain shapes, rhythms, feature types, etc. Here, we proposed a semi-supervised meta learning (SSML) algorithm to train EEG decoding model from source subjects and transfer it to new incoming target subjects.

\subsection{Model agnostic meta-learning algorithm}

Meta learning is to learn model for a variety of tasks and achieve best performance on a distribution of tasks, including potentially unseen tasks. In practical, the subject-transfer learning in BCI could be considered as a meta learning problem that each task refer to the classification of each subject‘s neural signals.

Meta learning approaches could be categorized into three kinds, i. e. model-based, metric-based and optimization-based \cite{Timothy2020}. Model-based method relies on designing suitable models for fast learning on new task. Metric-based approach aims to learn a good kernel of distance metric which is applicable to common tasks as well as the unseen tasks. Optimization-based method could adjust the optimization algorithm on new task after learning a few examples. Model-agnostic meta-learning (MAML) is a fairly general optimization algorithm, compatible with any model that learns through gradient descent \cite{finn_maml2017}. According to the recent studies \cite{boney2018, zhu2020, jeong2020, zhong2020, banluesombatkul2021}, MAML method achieved impressive performance in different kinds of few-shot learning problems.

Here is a dataset $D=<S, Q>$ where $S=\{(X_s, Y_s)\}$ is a support set of source subjects with EEG samples $X_s$ and the associated labels $Y_s$, and $Q=\{(X_q)\}$ is a query set of a target subject with unlabeled sample $X_q$. The subject-transfer learning problem in BCI is to learn a model with the support set and apply the model on target subject to predict the labels of query set. The model's parameter $\theta ^ *$ is optimized as follow:
\begin{equation}\label{inter}
{\theta ^ * } = arg\mathop {max}\limits_\theta  {\mathbb{E}_{(\boldsymbol{x},y) \in S}}[p_\theta({y}\left| \boldsymbol{x} \right.)]
\end{equation}
where $p$ denotes the posterior probability of the classifier to classify a given sample $\boldsymbol{x}$ to a specific class $y$ in support set $S$. By maximizing the expectation $\mathbb{E}$ of support set, the model is optimized. Furthermore, the optimized model could be applied on target subject to predict the labels of query set.

\makeatletter
\newcommand{\removelatexerror}{\let\@latex@error\@gobble}
\makeatother
\begin{figure}[!t]\label{Algorithm}
	\label{alg:LSB}
	\removelatexerror
 \begin{algorithm}[H]%
    \caption{Model agnostic meta-learning algorithm}%
	\begin{algorithmic}[1]
		\REQUIRE Backbone network $f_\theta$, base learning rate $\alpha$, meta learning rate $\beta$
        \REQUIRE Labeled data of source subjects $S=\{({X_s},{Y_s})\}_{s=1}^{N_s}$.
		\WHILE {not done}
			\STATE Sample a subset of $M$ subjects from source $S$.
    		\FOR {all $\mathcal{T}_i, i \in 1,...M$}
    		      \STATE Update base learner parameters $\theta^*$:
    		          \\$    \theta_i^* \leftarrow \theta - \alpha {\nabla _{\theta}}{\mathcal{L}_{\mathcal{T}_i}}({f_{\theta}})$
    		\ENDFOR
            \STATE Update meta-learner parameters $\theta$:
                        \\$\theta \leftarrow \theta - \beta {\nabla _{\theta}} \sum_{i=1}^{M}{\mathcal{L}_{\mathcal{T}_i}}({f_{\theta_i^*}})$
		\ENDWHILE
	\end{algorithmic}
\end{algorithm}
\end{figure}

\begin{figure*}[ht]
\centering
\includegraphics[width=0.99\textwidth]{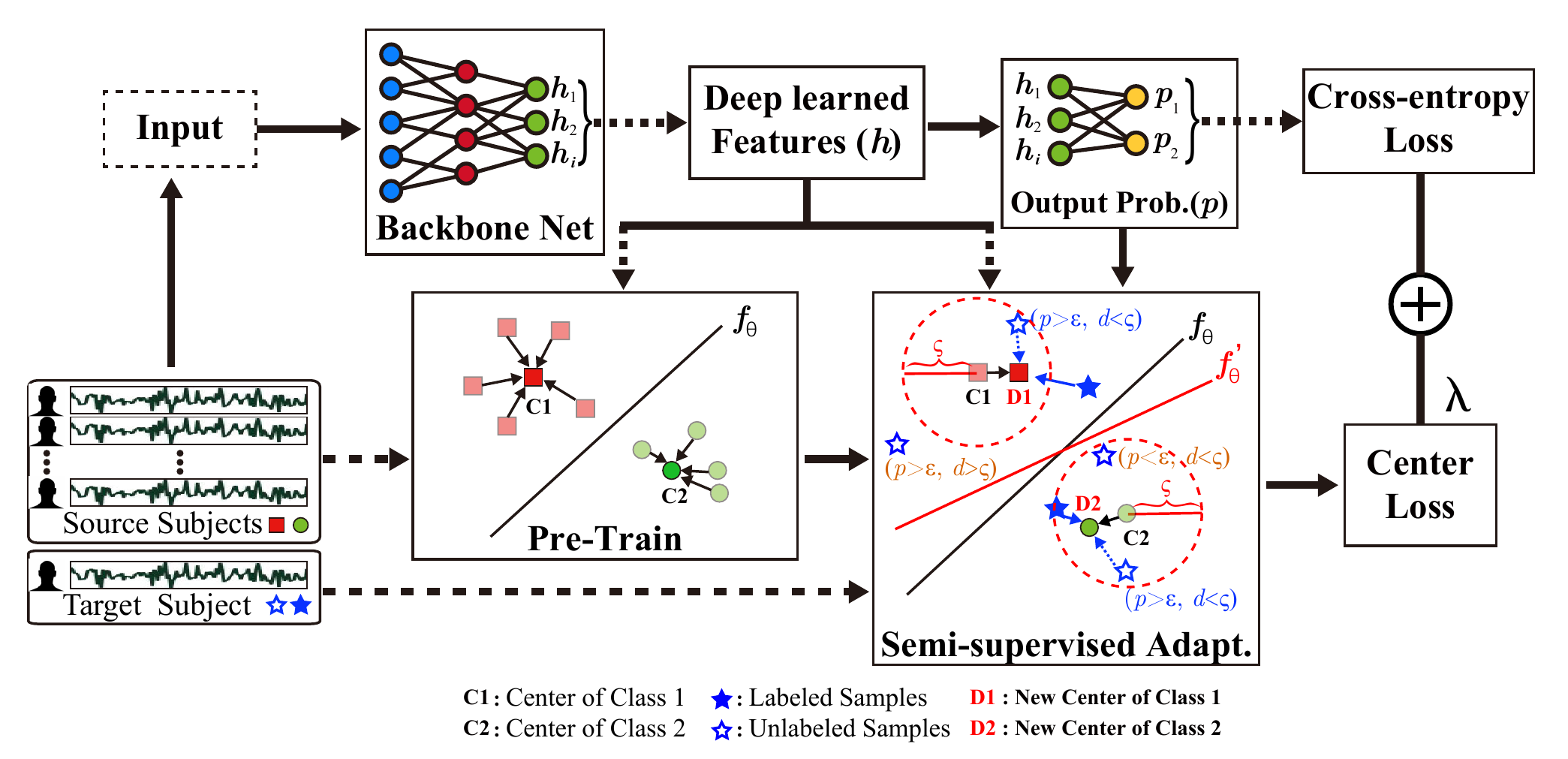}
\caption{Semi-supervised meta learning algorithm for subject-transfer EEG decoding.}
\label{afig0}
\end{figure*}

As shown in \textbf{Algorithm 1}, MAML algorithm is to learn a meta-learner by a support set with $N_s$ source subjects $S=\{({X_s},{Y_s})\}_{s=1}^{N_s}$. Given a model $f_{\theta}$ with parameters $\theta$, it could be optimized by gradient descent approach as
\begin{equation}\label{baselearn}
\theta _i^* \leftarrow \theta  - \alpha {\nabla _\theta }{\mathcal{L}_{{\mathcal{T}_i}}}({f_\theta })
\end{equation}
where $\mathcal{L}$ denotes the loss between the model's output probability and the associated label of the given data $\mathcal{T}_i$ of the $i$-th source subject, $\nabla _\theta$ is the gradient computed by the loss, $\alpha$ is the learning rate.

The above optimized parameters $\theta _i^*$ is just for one subject. For better generalization across the other subjects, we sample a subset of $M$ subjects from source $S$ and train a meta-learner for more efficient subject-transfer learning. The meta-learner's parameters could be optimized by the averaged gradients of multiple source subjects as
\begin{equation}\label{metalearn}
\theta \leftarrow \theta - \beta {\nabla _{\theta}} \sum\nolimits_{{\mathcal{T}_i}\in S}{\mathcal{L}_{{\mathcal{T}_i}}}({f_{\theta_i^*}})
\end{equation}
where $\mathcal{T}_i$ denotes the $i$-th subject in support set $S$, $\beta$ is the learning rate to train the meta-learner. For calibration on new subjects, the trained meta-learner is fine-tuned by few labeled samples of a new incoming subject with Equation (\ref{baselearn}).

According to the recent studies \cite{zhu2020, banluesombatkul2021, bontonou2021}, meta learning methods including MAML were used for subject-transfer EEG classification that they require some labeled data of the unseen subject for fast adaption. However, the EEG signals of a subject vary a lot in different sessions/time. Only few labeled samples are unlikely to capture the real distribution of EEG signals of a new incoming subject. Generally, the labeled data are scarce or expensive while unlabeled data are readily available in BCI applications and neuroscience studies. Therefore, we would like to combine semi-supervised learning and meta learning methods to transfer EEG decoding models for new subjects just with few-shot labeled and many unlabeled samples.

\subsection{Semi-supervised meta learning algorithm}

In the BCI applications like sleep staging, the labeled data are scarce or expensive and unlabeled data are readily available for new incoming subjects. Therefore, it is of great interests to develop a semi-supervised learning method which could exploit labeled and unlabeled data to fine-tune the model for target subject. According to the previous research, semi-supervised learning method was used for brain signal feature extraction \cite{singh2015, sun2019, xu2016affective} and improving signal decoding performance for EEG-based emotion recognition, sleep stage classification, diagnosis of brain disease, etc \cite{langkvist2012, chai2016, hsu2022, dan2021possibilistic}.

Here, we proposed a semi-supervised meta learning (SSML) algorithm for subject-transfer EEG decoding as shown in Fig. \ref{afig0}. The whole architecture could be divided into two phases, i. e. the pre-training phase and semi-supervised adaption phase.

In the pre-training phase, the backbone network $f_\theta$ is trained by the labeled EEG signals of source subjects. To enhance inter- and intra-subject discriminative features learned by deep backbone network, we applied an additional loss function (termed center loss) for regulation \cite{leibe2016}. The deep learned features are the outputs of the last hidden layer in the backbone network. The center loss function aims to learn a center for features of each class, and penalizes the distances between the features and the corresponding class centers. As a result, the center of each class \textbf{C1}/\textbf{C2} is pre-trained by source subjects as shown in Fig. \ref{afig0}.

In the semi-supervised adaption phase, the pre-trained network is fine-tuned by the averaged gradient of few-shot labeled and unlabeled samples of the target subject. Here, only the unlabeled samples with high output probabilities ($p>\varepsilon$) and close to the class center ($d<\varsigma$) are included for fine-tuning while the rest samples ($p<\varepsilon$ or $ d>\varsigma$) are ignored. Then, the previous class centers (\textbf{C1}, \textbf{C2}) are transferred to the new ones (\textbf{D1}, \textbf{D2}) for target subject that the network $f_\theta^{'}$ is fine-tuned for the target subject as shown in Fig. \ref{afig0}.

The joint loss function which combines center loss and cross entropy loss is as follows.
\begin{equation}\label{center}
\begin{array}{l}
{\mathcal{L}_C} = \frac{1}{2}\sum\limits_{i = 1}^m {\left\| {{\boldsymbol{h}_i} - {\boldsymbol{c}_{{y_i}}}} \right\|_2^2} \\
{\mathcal{L}_S} =  - \sum\limits_{i = 1}^m {{y_i}\log ({p_i})} \\
\mathcal{L} = {\mathcal{L}_S} + \lambda {\mathcal{L}_C}
\end{array}
\end{equation}
where $\boldsymbol{h}_i$ denotes the output feature of the last hidden layer of the backbone network given the $i$-th input sample, $\boldsymbol{c}_{y_i}$ is the $y_i$-th class center, $p_i$ is the output of the network, $y_i$ is the ground truth, ${\mathcal{L}_S}$ is cross-entropy classification loss, $\lambda$ is the weight of center loss ${\mathcal{L}_C}$. The feature center $\boldsymbol{c}_{y_i}$ for each class is considered as learnable parameters which is learned by gradient descent method \cite{leibe2016}. Its gradient is as follows.
\begin{equation}\label{centergradient}
\begin{array}{l}
\frac{{\partial {L_C}}}{{\partial {\boldsymbol{h}_i}}} = {\boldsymbol{h}_i} - {\boldsymbol{c}_{{y_i}}}\\
{\triangle{\boldsymbol{c}_j}} = \frac{{\sum\nolimits_{i = 1}^m {\delta ({y_i} = j) \cdot ({\boldsymbol{c}_j} - {\boldsymbol{h}_i})} }}{{1 + \sum\nolimits_{i = 1}^m {\delta ({y_i} = j)} }}
\end{array}
\end{equation}
where $\delta (condition) = 1$ if the condition is satisfied, and $\delta (condition) = 0$ if not.

\makeatletter
\makeatother
\begin{figure}[!t]\label{Algorithm2}
	\label{alg:LSB2}
	\removelatexerror
 \begin{algorithm}[H]%
    \caption{Semi-supervised meta learning algorithm}%
	\begin{algorithmic}[1]
		\REQUIRE Pre-trained network $f_\theta$, base learning rate $\alpha$, fine-tuning rate $\gamma$, confidence and distance thresholds $\varepsilon$, $\varsigma$.
        \REQUIRE Few-shot labeled signals $T_q=\{({\boldsymbol{x}_q},{y_q})\}_{q=1}^{N_q}$ and unlabeled signals ${X_t} = \{{\boldsymbol{x}_j}\}$ of target subject.
        \STATE Obtain output of network $(\boldsymbol{o}_j, \boldsymbol{h}_j) = f_{\theta}(\boldsymbol{x}_j), \boldsymbol{x}_j \in X_t$
        \STATE Generate artifact labels $(p_j, y_j) = arg\mathop {max(\boldsymbol{o}_j)}\limits_{} $
        \STATE Calculate distance from center ${d_j} = \frac{1}{2n}\left\| {{\boldsymbol{h}_j} - {\boldsymbol{c}_{{y_j}}}} \right\|_2^2$
        \STATE Construct support set: $Q=\{({\boldsymbol{x}_j},{y_j})\}, s.t. p_j>\varepsilon, d_j<\varsigma$
        \WHILE {not done} 
            \STATE Sample class-balanced support set $B=\{T_j\}$ from $Q$
            \FOR {all $T_j$}
                \STATE Update base learner parameters $\theta^*$:
		          \\$    \theta_j^* \leftarrow \theta - \alpha {\nabla _{\theta}}{\mathcal{L}_{{T_j}}}({f_{\theta}})$
            \ENDFOR
            \STATE Update meta-learner parameters $\theta$:
                        \\$\theta \leftarrow \theta - \gamma {\nabla _{\theta}} \sum_{T_j \in B}[{\mathcal{L}_{{T_j}}}({f_{\theta_j^*}}) +{\mathcal{L}_{{T_q}}}({f_{\theta_j^*}})]$
		\ENDWHILE
	\end{algorithmic}
\end{algorithm}
\end{figure}

Based on the Algorithm 1, the backbone network is pre-trained by the joint loss function as shown in Fig. \ref{afig0}. Then, the pre-trained network is fine-tuned in a semi-supervised manner with few-shot labeled EEG signals $T_q=\{({\boldsymbol{x}_q},{y_q})\}_{q=1}^{N_q}$ and unlabeled signals ${X_t} = \{{\boldsymbol{x}_j}\}$ of the target subject.

In the semi-supervised adaption phase, only part of the unlabeled signals are eligible and selected for fine-tuning. First, each unlabeled signal $\boldsymbol{x}_i \in X_t$ is fed into the network to obtain its feature $\boldsymbol{h}_j$ and the corresponding output $\boldsymbol{o}_j$ as shown in Algorithm 2. Second, the $argmax$ function is applied to transform the network output $\boldsymbol{o}_j$ into the artificial label $y_j$ and the associated predict probability $p_j$. Third, the feature distance of each signal from the corresponding class center is calculated. The feature distance denotes the element-wise averaged center loss of feature vector. Fourth, only the signals with high probabilities and close to the class center are included in the support set. The feature distance from center ${{d}_j}$ and the support set are defined as follows:
\begin{equation}\label{bls}
\begin{array}{l}
{{d}_j} = \frac{1}{2n}\left\| {{\boldsymbol{h}_j} - {\boldsymbol{c}_{{y_j}}}} \right\|_2^2 \\
Q=\{({\boldsymbol{x}_j},{y_j})\}, s.t. p_j>\varepsilon, {d}_j<\varsigma
\end{array}
\end{equation}
where $n$ is the length of each feature vector $\boldsymbol{h}_j$ for normalization of feature distance, $\varepsilon$ and $\varsigma$ are the confidence threshold and distance threshold, only the samples with high predicting probabilities ($p_j>\varepsilon$) and close to the class center ($d_j<\varsigma$) are included in the support set $Q$. Fifth, we down-sample and obtain a class-balanced subset $B$ from the support set for effective training and unbias classification on each class. Then, the backbone network is fine-tuned by the averaged gradient of the artificial support set $B$ and few-shot labeled samples $T_q$.

Consequently, the fine-tuned model could be applied for classification tasks of the target subject. With a semi-supervised adaption, the proposed SSML method utilizes labeled and unlabeled data to achieve better EEG decoding performance for new incoming target subject. Next, we will conduct a series of experiments to evaluate the proposed SSML method and compare it with the other methods.

\section{Experiments}

In this section, we present the details of the experiments on the proposed method. In addition, the corresponding results of the other state-of-the-art methods are also present for comparison. In our experiments, the hardware and software configuration used in our experiments are based on a platform with an Nvidia RTX 2080Ti, Ubuntu 16.04, PyTorch 1.9.0.

\subsection{Datasets}

In the experiments, the proposed method is verified by three different EEG paradigms including ERP detection, emotion recognition and sleep staging. The feasibility of the proposed method on different kinds of paradigms are significant.

ERP is a fundamental technique in BCI applications and neuroscience. The 32-channel EEG signals of four severely disabled and four able-bodied subjects were recorded and released as benchmark dataset for ERP detection \cite{hoffmann2008efficient}. The ERP/NoERP signals were extracted and averaged to enhance signal-to-noise rate. Then, there are 24 ERP and 120 NoERP signals for each subject. Each ERP/NoERP signal is represented by a $32*128$ matrix which denotes 32 channels and 128 time points in 1 second.

The SJTU emotion EEG dataset (SEED) is a benchmark dataset for evaluating EEG emotion recognition algorithms \cite{duan2013differential}. The 62-channel EEG signals stimulated by happy, sad and neutral film clips were recorded. We extracted the differential entropy features of 5 frequency channels in every 10 second frame that each sample is represented by a $62*10*5$ matrix. There are total 15 subjects and 993 samples for each subject in the dataset.

Sleep-EDF dataset from MCH-Westeinde Hospital was a benchmark dataset for EEG sleep staging methods \cite{Kemp2000}. Here, the sleep EEG signals in Fpz-Cz channel of 20 healthy subjects in Sleep-EDF dataset were used for experiments. Each EEG signal is segmented into 30-second frames, totally 3000 time-points in one frame. There are five sleep stages, i.e. wake(W), non-rapid eye movement (NREM: N1, N2, and N3), rapid eye movement (REM). The number of sleep EEG signals of each subject varies from 2000 to 6000.

Based on the meta learning methodology, the model could be pre-trained by labeled data of the existed source subjects and fine-tuned for a new incoming subject with only few labeled data. Therefore, leave-one-subject-out (LOSO) cross-validation strategy was applied for evaluating subject-transfer EEG decoding performance. For example, seven of the eight subjects in ERP dataset are selected as source subjects for training while the rest one is the target subject for testing. A few labeled samples of target subject are collected for fine-tuning the model. Then, the classification accuracies of the eight runs of LOSO experiments are averaged as final testing performance.

\subsection{Model}

As a variant of model-agnostic meta-learning, the proposed SSML method is also feasible on different network structures.
To study the performance variances of the proposed method on different network structures, we built three kinds of networks including multi-layer perception (MLP), spatial-temporal filtering neural network (STNN) and convolutional neural network (CNN).

In Table \ref{models}, three backbone models for ERP detection dataset are presented as examples. In the MLP model, the $32\times 128$ input signal is firstly flattened into a $1\times 4096$ vector and fed into a linear hidden layer with 300 Relu activation units and an output layer with 2 Softmax activation units. In the STNN model, a linear spatial filtering layer with 16 filters and a temporal filtering layer with 64 filters are used to extract spatial and temporal features, respectively. Then the obtained $16\times 64$ feature is flattened into a $1\times 1024$ vector and fed into the output layer. In the CNN model, convolutional (Conv) layers with Relu activation, Max-pooling (Pool) layers and a linear output layer with Softmax activation are used. As for emotion and sleep datasets, the backbone models are similar to those for ERP that needs to modify the structure to fit the signals with different number of channels or time lengths.
\begin{table*}[htbp]
  \centering
  \caption{Backbone models for ERP detection includes MLP, STNN and CNN.}
    \begin{tabular}{|clc|rr|clc|}
\cmidrule{1-3}\cmidrule{6-8}    \multicolumn{3}{|c|}{\textbf{MLP}} &       &       & \multicolumn{3}{c|}{\textbf{CNN}} \\
\cmidrule{1-3}\cmidrule{6-8}    \multicolumn{1}{|c|}{Layer} & \multicolumn{1}{c|}{Settings} & Output &       &       & Layer & \multicolumn{1}{c}{Settings} & Output \\
\cmidrule{1-3}\cmidrule{6-8}    Input &       & $32\times128$ &       &       & Input &       & $32\times128$ \\
\cmidrule{1-3}\cmidrule{6-8}    Flatten &       & $1\times4096$ &       &       & Conv  & ($1\times16$)*16 & (16, $32\times113$) \\
\cmidrule{1-3}\cmidrule{6-8}    Hidden & $4096\times300$ & $1\times300$ &       &       & Pool  & $1\times2$ & (16, $32\times56$) \\
\cmidrule{1-3}\cmidrule{6-8}    Output & $300\times2$ & 2     &       &       & Conv  & ($1\times3$)*32 & (32, $32\times54$) \\
\cmidrule{1-3}\cmidrule{6-8}    \multicolumn{1}{r}{} &       & \multicolumn{1}{r}{} &       &      & Pool  & $1\times2$ & (32, $32\times27$) \\
\cmidrule{1-3}\cmidrule{6-8}    \multicolumn{3}{|c|}{\textbf{STNN}} &       &       & Conv  & ($1\times3$)*64 & (64, $32\times25$) \\
\cmidrule{1-3}\cmidrule{6-8}    Layer & \multicolumn{1}{c}{Settings} & Output &       &       & Pool  & $1\times2$ & (64, $32\times12$) \\
\cmidrule{1-3}\cmidrule{6-8}    Input &       & $32\times128$ &       &       & Conv  & ($1\times3$)*128 & (128, $32\times10$) \\
\cmidrule{1-3}\cmidrule{6-8}    Spatial & \multicolumn{1}{c}{$16\times32$} & $16\times128$ &       &       & Conv  & \multicolumn{1}{c}{\multirow{2}[4]{*}{($32\times1$)*256}} & \multirow{2}[4]{*}{(256, $1\times10$)} \\
\cmidrule{1-3}    Temporal & \multicolumn{1}{c}{$128\times64$} & $16\times64$ &       &       & (Spatial) &       &  \\
\cmidrule{1-3}\cmidrule{6-8}    Flatten &       & $1\times1024$ &       &       & Flatten &       & $1\times2560$ \\
\cmidrule{1-3}\cmidrule{6-8}    Output & $1024\times2$ & 2     &       &       & Output & $2560\times2$ & 2 \\
\cmidrule{1-3}\cmidrule{6-8}    \end{tabular}%
  \label{models}%
\end{table*}%

To train the backbone network, we applied Adam optimizer with base learning rate of 0.001 and meta learning rate of 0.0001. To learn feature centers for different classes, stochastic gradient descent optimizer is applied with a learning rate of 0.001 and a center-loss weight of $0.001$. The confidence threshold and distance threshold are 0.9 and 1 which are used to generate support set for fine-tuning the backbone network.

In the pre-training phase, 90 percent of source subjects are used to train the backbone network while the rest 10 percent is the validation set for monitoring. Once the classification performance of the validation set is maximized, the training procedure will be stopped. For the target subject of LOSO experiment, few-shot labeled samples of each class are randomly selected for fine-tuning the pre-trained network. The network is fine-tuned by 10 epochs using Adam optimizer with learning rate of 0.001 and weight decay of 0.001.

\subsection{Results}

\begin{table}[htbp]
  \centering
  \caption{Performance of different methods on three EEG benchmark datasets.}
    \begin{tabular}{lrrlrrlr}
    \toprule
    \multicolumn{2}{c}{ERP Detection} &       & \multicolumn{2}{c}{Emotion Recognition} &       & \multicolumn{2}{c}{Sleep Staging} \\
\cmidrule{1-2}\cmidrule{4-5}\cmidrule{7-8}    Methods & \multicolumn{1}{c}{Accuracy} &       & \multicolumn{1}{c}{Method} & \multicolumn{1}{c}{Accuracy} &       & \multicolumn{1}{c}{Method} & \multicolumn{1}{c}{Accuracy} \\
\cmidrule{1-2}\cmidrule{4-5}\cmidrule{7-8}    BLDA \cite{hoffmann2008efficient} & 0.8511 &       & DGCNN \cite{song2019eeg} & 0.6639  &       & Stack SAE \cite{tsinalis2016automatic} & 0.7890  \\
\cmidrule{1-2}\cmidrule{4-5}\cmidrule{7-8}    STDA \cite{zhang2013spatial} & 0.8279 &       & DAN \cite{li2018cross} & 0.7134  &       & SVM \cite{liu2020diffuse} & 0.8272  \\
\cmidrule{1-2}\cmidrule{4-5}\cmidrule{7-8}    KNN \cite{alom2020classification} & 0.8564 &       & BiHDM \cite{li2020novel} & 0.7722  &       & CNN \cite{tsinalis2016automatic} & 0.7480  \\
\cmidrule{1-2}\cmidrule{4-5}\cmidrule{7-8}    SVM \cite{rakotomamonjy2008bci} & 0.8420 &       & RGNN \cite{zhong2020eeg} & 0.7957  &       & Multi-task CNN \cite{phan2018joint} & 0.8190  \\
\cmidrule{1-2}\cmidrule{4-5}\cmidrule{7-8}    CNN \cite{cecotti2010convolutional} & 0.8547 &       & SOGNN \cite{li2021cross} & 0.8104  &       & DeepSleepNet \cite{supratak2017deepsleepnet} & 0.8200  \\
    \midrule
    \midrule
W/O Meta  & 0.8329 &       & W/O Meta  & 0.7694  &       & W/O Meta  & 0.7942  \\
\cmidrule{1-2}\cmidrule{4-5}\cmidrule{7-8}    MAML \cite{finn_maml2017} & 0.8912 &       & MAML\cite{finn_maml2017} & 0.8355  &       & MAML\cite{finn_maml2017, banluesombatkul2021} & 0.8086  \\
\cmidrule{1-2}\cmidrule{4-5}\cmidrule{7-8}    \textbf{SSML(ours)} & \textbf{0.9521} &       & \textbf{SSML(ours)} & \textbf{0.8859}  &       & \textbf{SSML(ours)} & \textbf{0.8331}  \\
\cmidrule{1-2}\cmidrule{4-5}\cmidrule{7-8}    \end{tabular}%
  \label{results}%
\end{table}%

In Table \ref{results}, we present the performance of state-of-the-art methods and the proposed method on the benchmark datasets including ERP detection, emotion recognition and sleep staging tasks. According to the previous research \cite{banluesombatkul2021, bontonou2021}, MAML could be used for subject-transfer EEG classification, such as sleep staging. Here, we used convolutional neural network as backbone network for MAML method and the proposed SSML method.
In the MAML method, the labeled data of many source subjects were used to pre-train a backbone network by Algorithm 1. Then, the pre-trained backbone network will be fine-tuned by few labeled samples of target subject to achieve better performance.
The baseline method W/O Meta denotes the pre-trained network without using any data from target subject for fine-tuning.
In the ERP detection paradigm, we obtained the performance of the other methods by using sklearn toolbox or rebuilding the model with source subjects and labeled samples of target subject. For emotion recognition and sleep staging, we referred to the state-of-the-art performance recently.

In the proposed SSML method, the backbone network is pre-trained by the source subjects, then fine-tuned by few-shot labeled samples and unlabeled samples of target subject. For the emotion recognition and sleep staging paradigms, 10-shot labeled samples of each class are selected for fine-tuning in the MAML and SSML method. For the ERP paradigms with less samples of each subject, 5-shot setting is used in the experiments. The proposed SSML method achieved better performance than the other methods and MAML method in different EEG paradigms.

\begin{table}[htbp]
  \centering
  \caption{Classification performance improvements of MAML and SSML compared with W/O meta method.}
    \begin{tabular}{c|lrrrr}
    \toprule
    \multicolumn{1}{c}{Paradigm} & \multicolumn{1}{c}{Method} & \multicolumn{1}{c}{1-shot} & \multicolumn{1}{c}{3-shot} & \multicolumn{1}{c}{5-shot} & \multicolumn{1}{c}{10-shot} \\
    \midrule
    \multirow{2}[2]{*}{ERP} & MAML  & 4.80\% & 6.30\% & 7.00\% & 9.52\% \\
          & \textbf{SSML} & \textbf{10.00\%} & \textbf{12.60\%} & \textbf{14.31\%} & \textbf{16.10\%} \\
    \midrule
    \multirow{2}[2]{*}{Emotion} & MAML  & 4.40\% & 4.00\% & 6.70\% & 8.59\% \\
          & \textbf{SSML} & \textbf{5.90\%} & \textbf{10.90\%} & \textbf{10.30\%} & \textbf{15.14\%} \\
    \midrule
    \multirow{2}[2]{*}{Sleep} & MAML  & 0.80\% & 1.50\% & 1.60\% & 1.80\% \\
          & \textbf{SSML} & \textbf{2.20\%} & \textbf{3.80\%} & \textbf{3.40\%} & \textbf{4.90\%} \\
    \bottomrule
    \end{tabular}%
  \label{improve}%
\end{table}%

Moreover, we would like to study how much the classification performances are improved when using few labeled samples for fine-tuning with the MAML and the proposed SSML methods. The classification performance of the W/O meta method is considered as the baseline performance for each paradigm. Compared with the baseline, the SSML method achieved great improvements with only 1-shot labeled sample of each class as shown in Table \ref{improve}. As the increase of labeled samples, the MAML and SSML methods achieved much higher performance. That is to say, using more labeled samples is helpful to model real distribution of the target subject's EEG signals that the EEG decoding performance could be substantially improved. Consequently, the proposed SSML method with 10-shot labeled samples achieved over $15\%$ improvement in the ERP and emotion recognition paradigms and $4.9\%$ in sleep staging paradigm. Moreover, the proposed SSML method achieved higher improvements than MAML in different few-shot settings.

\begin{table}[htbp]
  \centering
  \caption{Wilcoxon signed rank test between SSML and the baseline methods.}
    \begin{threeparttable}
    \begin{tabular}{c|lcccc}
    \toprule
    \multicolumn{1}{c}{Paradigm} & \multicolumn{1}{c}{SSML VS \#} & 1-shot & 3-shot & 5-shot & 10-shot \\
    \midrule
    \multirow{2}[2]{*}{ERP} & W/O Meta & $\dag \dag$    & $\dag \dag$    & $\dag \dag$    & $\dag \dag$ \\
          & MAML  & $\dag$     & $\dag$     & $\dag \dag$    & $\dag \dag$ \\
    \midrule
    \multirow{2}[2]{*}{Emotion} & W/O Meta & $\dag \dag$    & $\dag$     & $\dag \dag$    & $\dag \dag$ \\
          & MAML  & $\dag$     & $\dag \dag$    & $\dag$     & $\dag \dag$ \\
    \midrule
    \multirow{2}[2]{*}{Sleep} & W/O Meta & $\sim$     &  $\dag$   & $\dag \dag$    & $\dag \dag$ \\
          & MAML  & $\sim$     &  $\sim$     & $\dag \dag$    & $\dag \dag$ \\
    \bottomrule
    \end{tabular}%
    \begin{tablenotes}
    \item[] Note: $\sim$ nonsignificant, $\dag p<0.05$, $\dag \dag p<0.01$
    \end{tablenotes}
    \end{threeparttable}
  \label{statistical}%
\end{table}%

Furthermore, we wonder whether the performance improvements of the proposed SSML method are statistically significant. Therefore, the performance differences between the proposed method and the other methods should be studied by statistical analysis method. As shown in Table \ref{statistical}, the Wilcoxon signed rank test results between the proposed SSML and the W/O meta method or MAML method are presented. In the ERP detection and Emotion recognition paradigms, the proposed SSML method achieved significant higher performance than the other two methods in different few-shot settings. In the sleep staging paradigm, significant improvements are found in 5-shot and 10-shot settings. The experimental results demonstrated the effectiveness and potential of the proposed SSML method in improving subject-transfer EEG decoding performance.

\section{Discussion}

Here, the proposed method is discussed in qualitative and quantitative manner.
First, to analyse the convergence procedure of different methods with different backbone networks on different EEG paradigms, the learning curves are present. Second, we discuss the subject-transfer decoding performance when changing the number of labeled samples of target subject. Third, we change the weight of center loss and analyse its contribution for improving classification performance.
Finally, we utilized the t-SNE technique to analyse how the proposed SSML method influences the classification in the feature domain.

\subsection{Comparison of Learning Curves}

\begin{figure}[ht]
\centering
\includegraphics[width=0.99\textwidth]{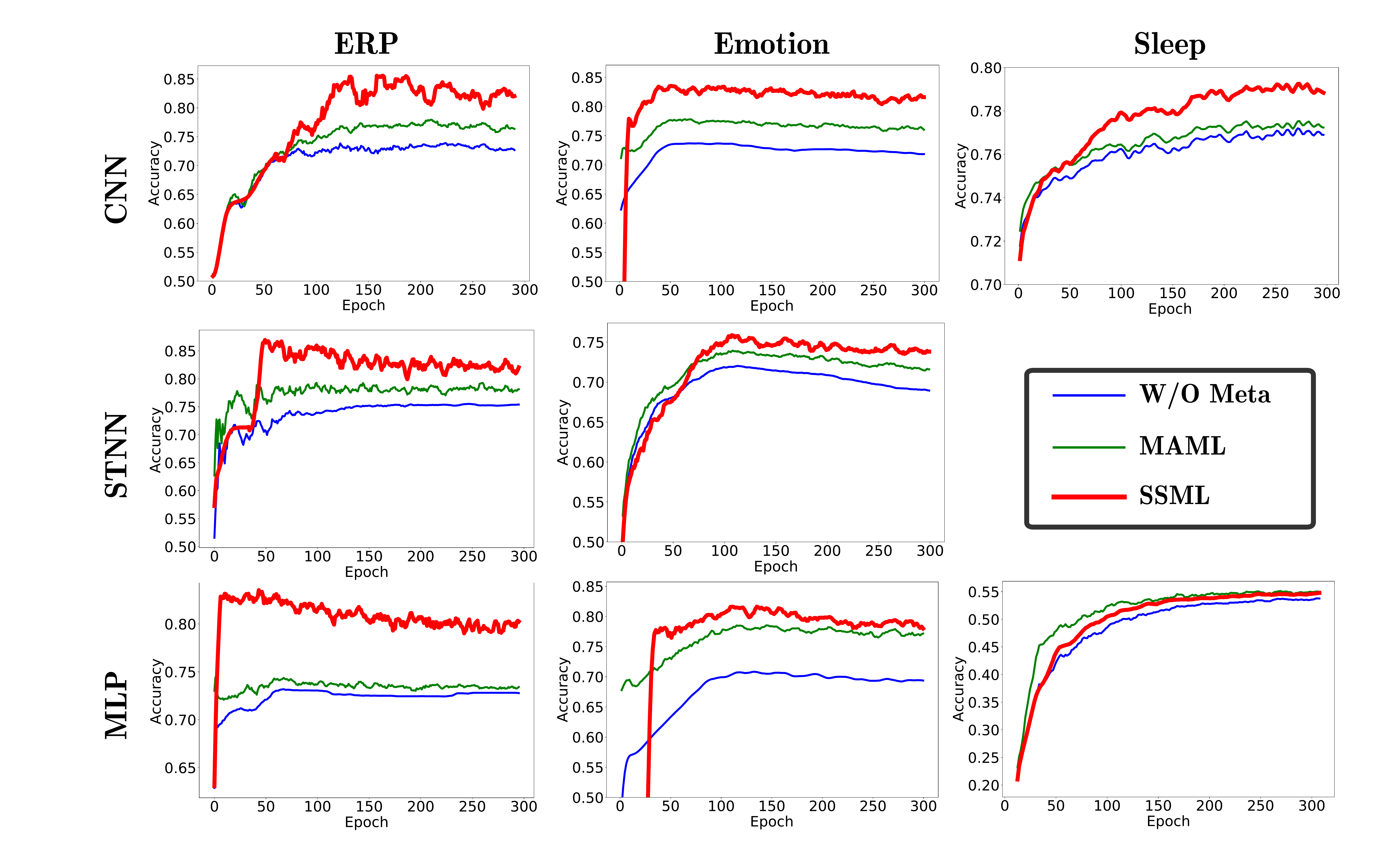}
\caption{Learning curves of three methods with different networks on different EEG paradigms. }
\label{afig1}
\end{figure}

Three different backbone models are trained by different meta-learning methods and evaluated on three datasets. The convergence procedure of different methods could be visualized by their learning curves as the increase of iteration epochs \cite{DBLP2015}. In Fig. \ref{afig1}, the learning curves of three different models trained by different meta-learning methods on three EEG datasets are presented. Here, each learning curve denotes the averaged classification accuracy in the LOSO cross-validation experiments.

Firstly, we evaluated three different kinds of neural networks including MLP, STNN and CNN. The signal of sleep dataset is one-channel EEG signal that the STNN with spatial filtering is infeasible and ignored. Apparently, the CNN achieved the higher performance than MLP or STNN. That is to say, the network structure also plays an important role in subject-transfer EEG decoding.
Secondly, we could find that the proposed SSML method achieved better performance than the other meta-learning methods. Specifically, its learning curve (red) is higher than the other methods within the same model and dataset.
Thirdly, the SSML method could achieve faster convergence than the other two methods in the experiments of three different networks.
The comparison of learning curves demonstrated the effectiveness of the SSML with CNN network in different paradigms.

\subsection{Consideration of N-shot Learning}

The calibration procedure of BCI for new subjects could be considered as few-shot learning problem. In the meta learning methods, few labeled samples of target subject are obtained for calibration, i.e. fine-tuning the EEG decoding model for this subject. In the SSML method, few labeled samples and many unlabeled samples are jointly used for fine-tuning.

\begin{figure}[ht]
\centering
\includegraphics[width=0.75\textwidth]{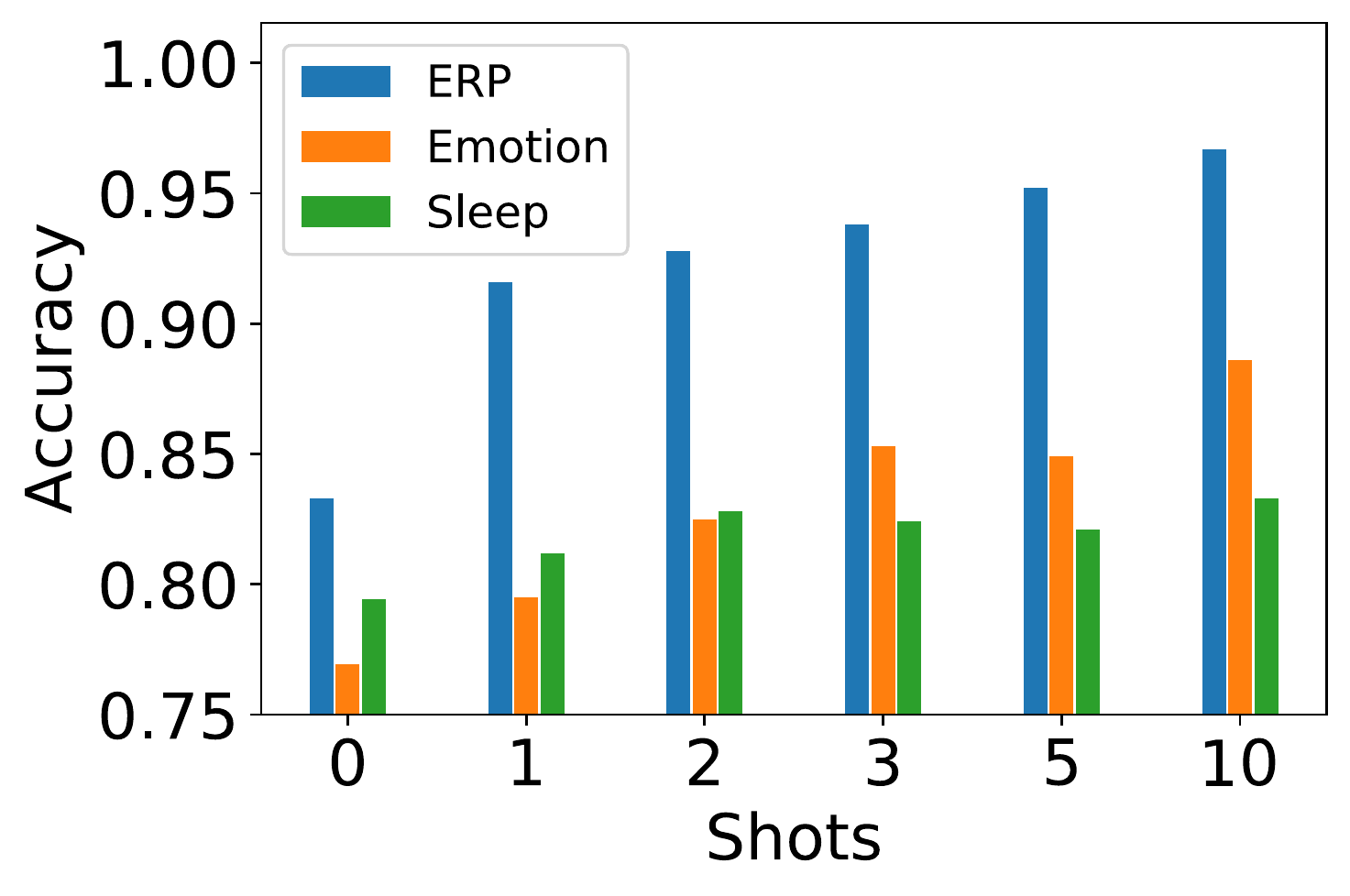}
\caption{Performance of SSML method with different shots of labeled samples.}
\label{afig3}
\end{figure}

Here, we would like to analyse the influence of the number of labeled samples for calibration. In Fig. \ref{afig3}, the classification accuracies of the SSML method with 0, 1, 2, 3, 5 and 10-shot settings are presented. The SSML in 0-shot setting denotes that only unlabeled samples are used for fine-tuning for target subject. Without labeled data of target subject, the classification performance is relatively poor. As the increase of the number of shots, the EEG decoding model is calibrated more accurately for the target subject. As a result, the classification accuracies are improved, especially in the ERP and emotion paradigms. The improved classification performance demonstrated the effectiveness of the proposed SSML for few-shot learning and calibration on new subjects.

\subsection{Effect of Center Loss weights}

\begin{figure}[ht]
\centering
\includegraphics[width=0.75\textwidth]{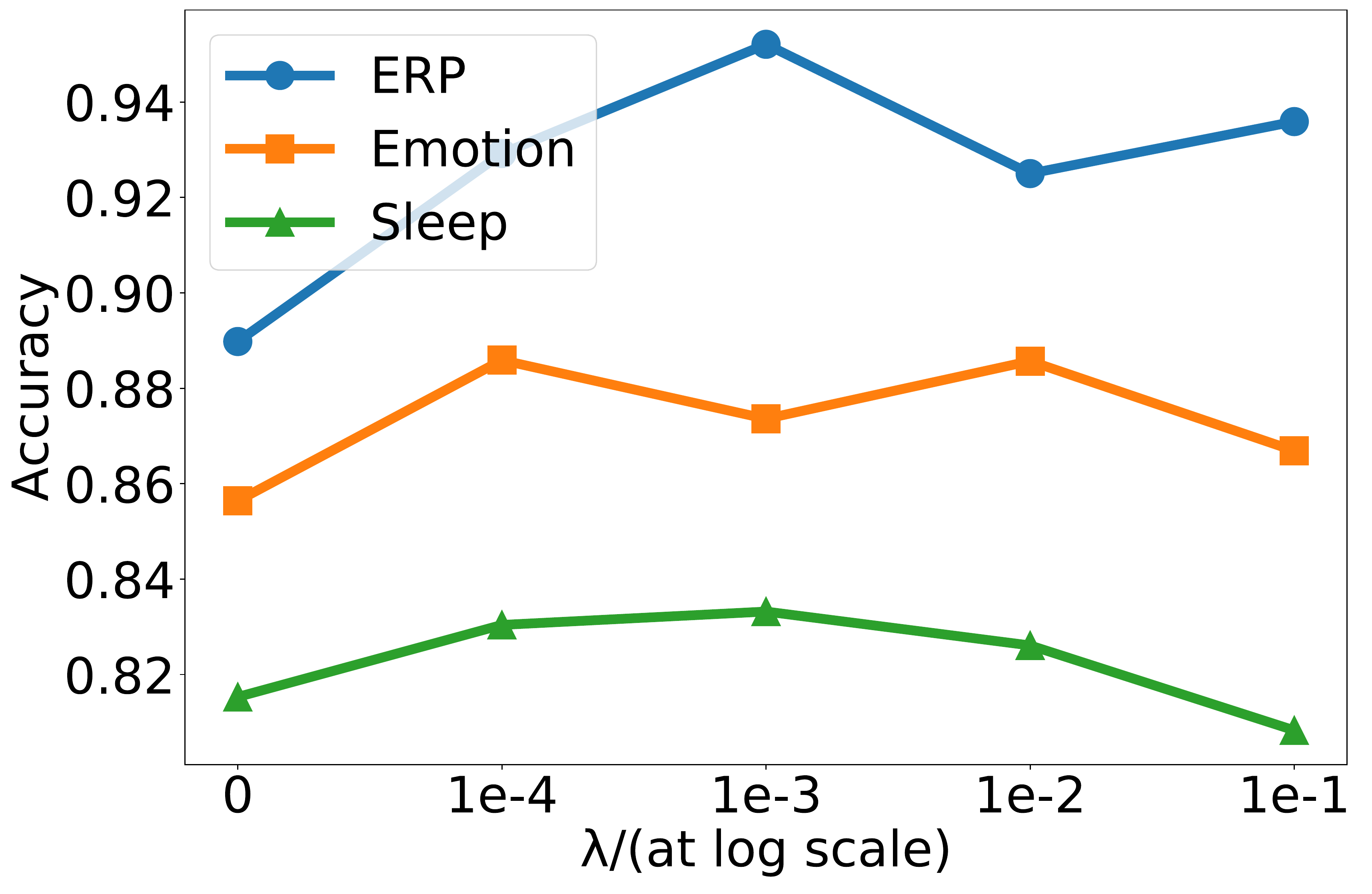}
\caption{Performance of SSML method with different center weights.}
\label{afig4}
\end{figure}

As aforementioned, the center loss and the associated feature distance are essential in the pre-training phase and semi-supervised adaption phase of SSML. Here, we would like to analyse the contribution and influence of center loss in decoding different EEG signals. The center loss function plays important role in dominating the intra-class variations in the joint loss function. It is intuitive to compare the classification performance when using joint loss function with different center weights. As shown in Fig. \ref{afig4}, the center weights between $1e-4$ and $1e-2$ have higher performance than the other weights that is similar to the previous research \cite{leibe2016}. The center weight of $1e-3$ is used in our experiments.


\subsection{T-SNE Analysis}

\begin{figure}[ht]
\centering
\includegraphics[width=0.75\textwidth]{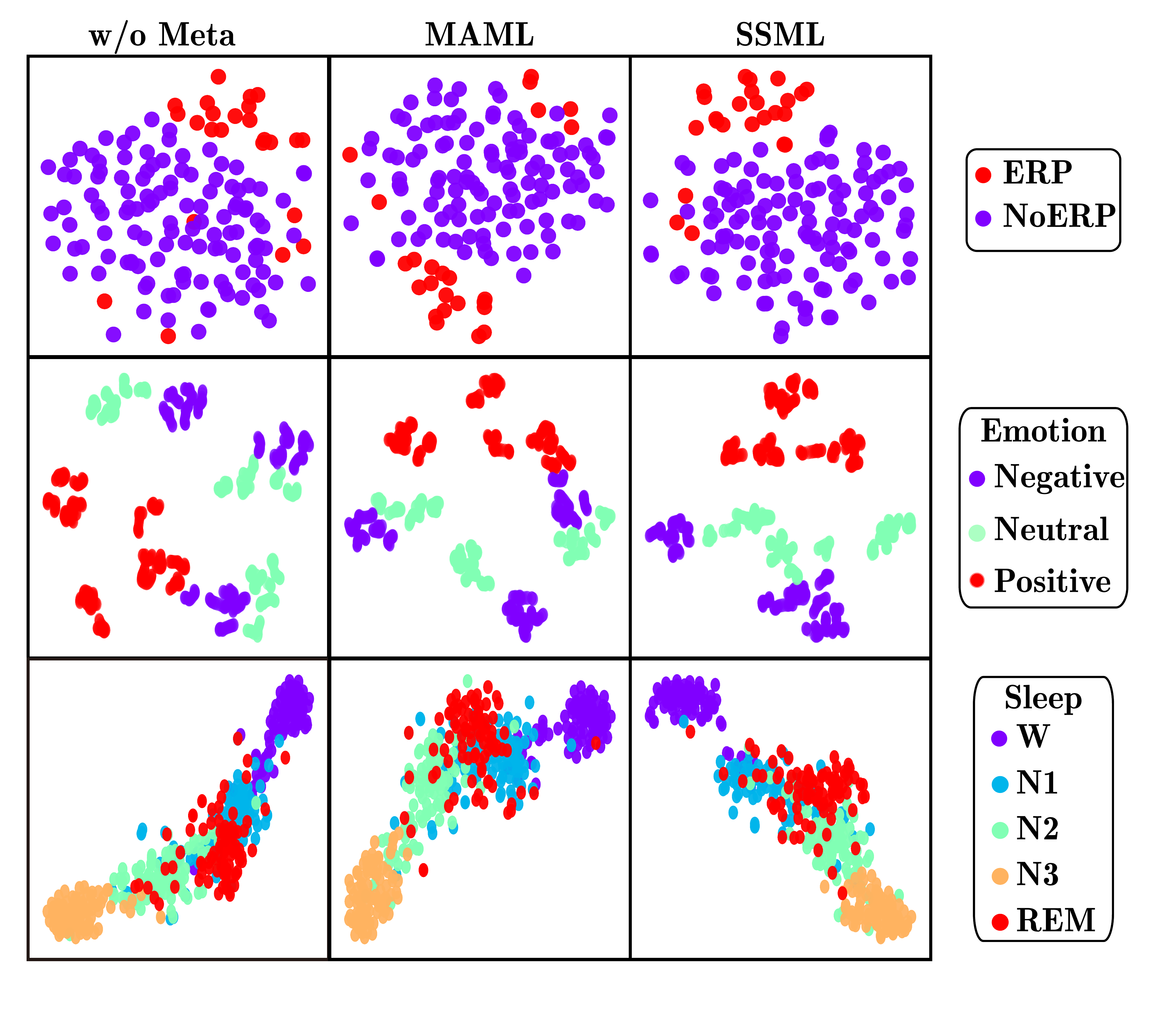}
\caption{T-SNE results of different methods in different paradigms. }
\label{afig2}
\end{figure}

As shown in the Fig. \ref{afig0}, the EEG features of different subjects are learned by the backbone network. We would like to analyse the differences of features obtained by different methods. To visualize high-dimensional features, t-distributed stochastic neighbor embedding (t-SNE) method is widely used. In Fig. \ref{afig2}, the t-SNE results of features learned by different method with CNN as backbone network are presented. Here, we chose one subject from each dataset for presentation. In each subfigure, the t-SNE results are presented in scatterplot in which the different colors denote the labels of samples. The samples of ERP and NoERP, three emotion classes, five sleep stages are presented.

In the t-SNE maps of the W/O meta method, the points from different classes are mixed with each other that the classification performance is not well. With MAML, a few labeled samples are used for fast adaption that the pre-trained network could be fine-tuned for the target subject. Based on the proposed SSML method, few labeled and many unlabeled samples are combined for semi-supervised adaption. Compared with the other methods, the samples of ERP class, three emotion class and wake (W) stage are likely separated by the SSML method. The t-SNE results verified that the proposed SSML method achieved better subject-transfer EEG decoding performance.

\subsection{Limitations}
The limitations of this study are as follows. First, the proposed method is only verified by three EEG paradigms with three backbone network structures in the experiments. Further researches on the other kinds of paradigms and networks should be conducted to verify the feasibility of the proposed method. Second, given a backbone network with relatively low performance, its performance will be even worse after apply semi-supervised adaption. That is to say, the proposed SSML method depends heavily on the effective network structures. Third, the proposed SSML method will be invalid if the features of the target subject are completely different with source subjects. Because the semi-supervised adaption phase of SSML is based on the backbone network pre-trained by the source subjects. Fourth, it also takes much times to collect few-shot labeled and many unlabeled samples for the calibration procedure of the proposed method. Further work is thus required to improve subject-transfer and calibration methods for EEG decoding in BCI as well as neuroscience.

\section{Conclusion}

In this paper, we propose a semi-supervised meta learning (SSML) method for subject-transfer EEG decoding in brain-computer interfaces. In a semi-supervised manner, the proposed method could effectively transfer the model trained by source subjects to new incoming subject. To verify the method, three different EEG paradigms including event-related potential, emotion recognition and sleep staging are tested. By using few-shot labeled and many unlabeled samples for fine-tuning, the proposed method could obviate inter- and intra-subject variability between source subjects and target subject. With a semi-supervised adaption procedure, the SSML achieved significantly improvements of target subjects. Moreover, the proposed method achieved better performance than the existing methods. Therefore, the proposed semi-supervised meta-learning method has much potentials in improving EEG decoding performance which is significant for practical BCI applications as well as neuroscience research.

\bibliography{ssml}

\begin{thebibliography}{10}
\expandafter\ifx\csname url\endcsname\relax
  \def\url#1{\texttt{#1}}\fi
\expandafter\ifx\csname urlprefix\endcsname\relax\def\urlprefix{URL }\fi
\expandafter\ifx\csname href\endcsname\relax
  \def\href#1#2{#2} \def\path#1{#1}\fi

\bibitem{lotte2015signal}
F.~Lotte, Signal processing approaches to minimize or suppress calibration time
  in oscillatory activity-based brain--computer interfaces, Proceedings of the
  IEEE 103~(6) (2015) 871--890.

\bibitem{hoffmann2008efficient}
U.~Hoffmann, J.-M. Vesin, T.~Ebrahimi, K.~Diserens, An efficient p300-based
  brain--computer interface for disabled subjects, Journal of Neuroscience
  methods 167~(1) (2008) 115--125.

\bibitem{lijp2019}
J.~Li, S.~Qiu, Y.-Y. Shen, C.-L. Liu, H.~He, Multisource {Transfer} {Learning}
  for {Cross}-{Subject} {{EEG}} {Emotion} {Recognition}, IEEE Transactions on
  Cybernetics (2019) 1--13.

\bibitem{banluesombatkul2021}
N.~Banluesombatkul, P.~Ouppaphan, P.~Leelaarporn, P.~Lakhan, B.~Chaitusaney,
  N.~Jaimchariyatam, E.~Chuangsuwanich, W.~Chen, H.~Phan, N.~Dilokthanakul,
  T.~Wilaiprasitporn, {MetaSleepLearner}: {A} {Pilot} {Study} on {Fast}
  {Adaptation} of {Bio}-{Signals}-{Based} {Sleep} {Stage} {Classifier} to {New}
  {Individual} {Subject} {Using} {Meta}-{Learning}, IEEE Journal of Biomedical
  and Health Informatics 25~(6) (2021) 1949--1963.

\bibitem{muhl2014survey}
C.~M{\"u}hl, B.~Allison, A.~Nijholt, G.~Chanel, A survey of affective brain
  computer interfaces: principles, state-of-the-art, and challenges,
  Brain-Computer Interfaces 1~(2) (2014) 66--84.

\bibitem{duan2013differential}
R.-N. Duan, J.-Y. Zhu, B.-L. Lu, Differential entropy feature for {EEG}-based
  emotion classification, in: 2013 6th International IEEE/EMBS Conference on
  Neural Engineering (NER), 2013, pp. 81--84.

\bibitem{chriskos2021review}
P.~Chriskos, C.~A. Frantzidis, C.~M. Nday, P.~T. Gkivogkli, P.~D. Bamidis,
  C.~Kourtidou-Papadeli, A review on current trends in automatic sleep staging
  through bio-signal recordings and future challenges, Sleep medicine reviews
  55 (2021) 101377.

\bibitem{Kemp2000}
B.~Kemp, A.~Zwinderman, B.~Tuk, H.~Kamphuisen, J.~Oberye, Analysis of a
  sleep-dependent neuronal feedback loop: the slow-wave microcontinuity of the
  {EEG}, IEEE Transactions on Biomedical Engineering 47~(9) (2000) 1185--1194.

\bibitem{kieffaber2016evaluation}
P.~D. Kieffaber, H.~R. Okhravi, J.~N. Hershaw, E.~C. Cunningham, Evaluation of
  a clinically practical, erp-based neurometric battery: application to
  age-related changes in brain function, Clinical Neurophysiology 127~(5)
  (2016) 2192--2199.

\bibitem{wei2018subject}
C.-S. Wei, Y.-P. Lin, Y.-T. Wang, C.-T. Lin, T.-P. Jung, A subject-transfer
  framework for obviating inter-and intra-subject variability in {EEG}-based
  drowsiness detection, NeuroImage 174 (2018) 407--419.

\bibitem{kwon2020}
O.-Y. Kwon, M.-H. Lee, C.~Guan, S.-W. Lee, Subject-{Independent}
  {Brain}¨c{Computer} {Interfaces} {Based} on {Deep} {Convolutional} {Neural}
  {Networks}, IEEE Transactions on Neural Networks and Learning Systems 31~(10)
  (2020) 3839--3852.

\bibitem{Wu2020}
D.~Wu, Y.~Xu, B.-L. Lu, Transfer learning for {EEG}-based brain-computer
  interfaces: A review of progress made since 2016, IEEE Transactions on
  Cognitive and Developmental Systems (2020) 1--1.

\bibitem{saha2020intra}
S.~Saha, M.~Baumert, Intra-and inter-subject variability in {EEG}-based
  sensorimotor brain computer interface: a review, Frontiers in computational
  neuroscience 13 (2020) 87.

\bibitem{congedo2017}
M.~Congedo, A.~Barachant, R.~Bhatia, Riemannian geometry for {{EEG}}-based
  brain-computer interfaces; a primer and a review, Brain-Computer Interfaces
  4~(3) (2017) 155--174, number: 3.

\bibitem{zancanaro2021cnn}
A.~Zancanaro, G.~Cisotto, J.~R. Paulo, G.~Pires, U.~J. Nunes, {CNN}-based
  approaches for cross-subject classification in motor imagery: From the
  state-of-the-art to dynamicnet, arXiv preprint arXiv:2105.07917.

\bibitem{CHEN2022}
C.~Chen, C.-M. Vong, S.~Wang, H.~Wang, M.~Pang, Easy domain adaptation for
  cross-subject multi-view emotion recognition, Knowledge-Based Systems 239
  (2022) 107982.

\bibitem{Ning2021}
R.~Ning, C.~Philip~Chen, T.~Zhang, Cross-subject {EEG} emotion recognition
  using domain adaptive few-shot learning networks, in: IEEE International
  Conference on Bioinformatics and Biomedicine (BIBM), 2021, pp. 1468--1472.

\bibitem{Li2021}
D.~Li, P.~Ortega, X.~Wei, A.~Faisal, Model-agnostic meta-learning for {EEG}
  motor imagery decoding in brain-computer-interfacing, in: IEEE/EMBS
  Conference on Neural Engineering (NER), 2021, pp. 527--530.

\bibitem{Choi2021}
S.~Choi, Meta-learning: Towards fast adaptation in multi-subject {EEG}
  classification, in: 2021 9th International Winter Conference on
  Brain-Computer Interface ({BCI}), 2021, pp. 1--1.

\bibitem{duan2020ultra}
T.~Duan, M.~Chauhan, M.~A. Shaikh, J.~Chu, S.~Srihari, Ultra efficient transfer
  learning with meta update for cross subject {EEG} classification, arXiv
  preprint arXiv:2003.06113.

\bibitem{choi2021meta}
S.~Choi, Meta-learning: Towards fast adaptation in multi-subject {EEG}
  classification, in: 2021 9th International Winter Conference on
  Brain-Computer Interface ({BCI}), IEEE, 2021, pp. 1--1.

\bibitem{bhosale2022calibration}
S.~Bhosale, R.~Chakraborty, S.~K. Kopparapu, Calibration free meta learning
  based approach for subject independent {EEG} emotion recognition, Biomedical
  Signal Processing and Control 72 (2022) 103289.

\bibitem{an2020}
S.~An, S.~Kim, P.~Chikontwe, S.~H. Park, Few-shot relation learning with
  attention for {{EEG}}-based motor imagery classification, in: 2020
  {IEEE}/{RSJ} International Conference on Intelligent Robots and Systems
  ({IROS}), {IEEE}, pp. 10933--10938.

\bibitem{wang2021}
Y.~Wang, S.~Qiu, X.~Ma, H.~He, A prototype-based {SPD} matrix network for
  domain adaptation {{EEG}} emotion recognition, Pattern Recognition 110 (2021)
  107626.

\bibitem{CHAKLADAR202117}
D.~D. Chakladar, P.~Kumar, P.~P. Roy, D.~P. Dogra, E.~Scheme, V.~Chang, A
  multimodal-siamese neural network (msnn) for person verification using
  signatures and {EEG}, Information Fusion 71 (2021) 17--27.

\bibitem{finn_maml2017}
C.~Finn, P.~Abbeel, S.~Levine, Model-{Agnostic} {Meta}-{Learning} for {Fast}
  {Adaptation} of {Deep} {Networks}, arXivArXiv: 1703.03400.

\bibitem{bontonou2021}
M.~Bontonou, G.~Lioi, N.~Farrugia, V.~Gripon, Few-{Shot} {Decoding} of {Brain}
  {Activation} {Maps}, in: 2021 29th {European} {Signal} {Processing}
  {Conference} ({EUSIPCO}), IEEE, Dublin, Ireland, 2021, pp. 1326--1330.

\bibitem{Timothy2020}
T.~M. Hospedales, A.~Antoniou, P.~Micaelli, A.~J. Storkey, Meta-learning in
  neural networks: {A} survey, CoRR abs/2004.05439.
\newblock \href {http://arxiv.org/abs/2004.05439} {\path{arXiv:2004.05439}}.

\bibitem{boney2018}
R.~Boney, A.~Ilin, Semi-supervised few-shot learning with maml (2018) 4.

\bibitem{zhu2020}
Y.~Zhu, M.~Saqib, E.~Ham, S.~Belhareth, R.~Hoffman, M.~D. Wang, Mitigating
  {Patient}-to-{Patient} {Variation} in {{EEG}} {Seizure} {Detection} using
  {Meta} {Transfer} {Learning}, in: {IEEE} {International} {Conference} on
  {Bioinformatics} and {Bioengineering} ({BIBE}), IEEE, 2020, pp. 548--555.

\bibitem{jeong2020}
T.~Jeong, H.~Kim, {OOD}-{MAML}: {Meta}-{Learning} for {Few}-{Shot}
  {Out}-of-{Distribution} {Detection} and {Classification}, in: H.~Larochelle,
  M.~Ranzato, R.~Hadsell, M.~F. Balcan, H.~Lin (Eds.), Advances in {Neural}
  {Information} {Processing} {Systems}, Vol.~33, Curran Associates, Inc., 2020,
  pp. 3907--3916.

\bibitem{zhong2020}
P.~Zhong, D.~Wang, C.~Miao, {{EEG}}-{Based} {Emotion} {Recognition} {Using}
  {Regularized} {Graph} {Neural} {Networks}, IEEE Transactions on Affective
  Computing (2020) 1--1.

\bibitem{singh2015}
G.~Singh, L.~Samavedham, Unsupervised learning based feature extraction for
  differential diagnosis of neurodegenerative diseases: {A} case study on
  early-stage diagnosis of {Parkinson} disease, Journal of Neuroscience Methods
  256 (2015) 30--40.

\bibitem{sun2019}
L.~Sun, B.~Jin, H.~Yang, J.~Tong, C.~Liu, H.~Xiong, Unsupervised {{EEG}}
  feature extraction based on echo state network, Information Sciences 475
  (2019) 1--17.

\bibitem{xu2016affective}
H.~Xu, K.~N. Plataniotis, Affective states classification using {EEG} and
  semi-supervised deep learning approaches, in: IEEE International Workshop on
  Multimedia Signal Processing (MMSP), IEEE, 2016, pp. 1--6.

\bibitem{langkvist2012}
M.~Langkvist, L.~Karlsson, A.~Loutfi, Sleep {Stage} {Classification} {Using}
  {Unsupervised} {Feature} {Learning}, Adv. Artif. Neu. Sys. 2012 (2012) 5.

\bibitem{chai2016}
X.~Chai, Q.~Wang, Y.~Zhao, X.~Liu, O.~Bai, Y.~Li, Unsupervised domain
  adaptation techniques based on auto-encoder for non-stationary {{EEG}}-based
  emotion recognition, Computers in Biology and Medicine 79 (2016) 205--214.

\bibitem{hsu2022}
S.-H. Hsu, Y.~Lin, J.~Onton, T.-P. Jung, S.~Makeig, Unsupervised learning of
  brain state dynamics during emotion imagination using high-density {{EEG}},
  NeuroImage 249 (2022) 118873.

\bibitem{dan2021possibilistic}
Y.~Dan, J.~Tao, J.~Fu, D.~Zhou, Possibilistic clustering-promoting
  semi-supervised learning for {EEG}-based emotion recognition, Frontiers in
  Neuroscience 15.

\bibitem{leibe2016}
Y.~Wen, K.~Zhang, Z.~Li, Y.~Qiao, ({Center} loss) {A} {Discriminative}
  {Feature} {Learning} {Approach} for {Deep} {Face} {Recognition}, in: Computer
  {Vision} ¨C {ECCV} 2016, Vol. 9911, 2016, pp. 499--515.

\bibitem{song2019eeg}
T.~{Song}, W.~{Zheng}, P.~{Song}, Z.~{Cui}, {EEG} emotion recognition using
  dynamical graph convolutional neural networks, IEEE Transactions on Affective
  Computing 11~(3) (2018) 532--541.

\bibitem{tsinalis2016automatic}
O.~Tsinalis, P.~M. Matthews, Y.~Guo, Automatic sleep stage scoring using
  time-frequency analysis and stacked sparse autoencoders, Annals of biomedical
  engineering 44~(5) (2016) 1587--1597.

\bibitem{zhang2013spatial}
Y.~Zhang, G.~Zhou, Q.~Zhao, J.~Jin, X.~Wang, A.~Cichocki, Spatial-temporal
  discriminant analysis for erp-based brain-computer interface, IEEE
  Transactions on Neural Systems and Rehabilitation Engineering 21~(2) (2013)
  233--243.

\bibitem{li2018cross}
H.~Li, Y.-M. Jin, W.-L. Zheng, B.-L. Lu, Cross-subject emotion recognition
  using deep adaptation networks, in: International conference on neural
  information processing, Springer, 2018, pp. 403--413.

\bibitem{liu2020diffuse}
G.-R. Liu, Y.-L. Lo, J.~Malik, Y.-C. Sheu, H.-T. Wu, Diffuse to fuse {EEG}
  spectra--intrinsic geometry of sleep dynamics for classification, Biomedical
  Signal Processing and Control 55 (2020) 101576.

\bibitem{alom2020classification}
M.~K. Alom, S.~M.~R. Islam, Classification for the p300-based brain computer
  interface ({BCI}), in: International Conference on Advanced Information and
  Communication Technology (ICAICT), IEEE, 2020, pp. 387--391.

\bibitem{li2020novel}
Y.~Li, L.~Wang, W.~Zheng, Y.~Zong, L.~Qi, Z.~Cui, T.~Zhang, T.~Song, A novel
  bi-hemispheric discrepancy model for {EEG} emotion recognition, IEEE
  Transactions on Cognitive and Developmental Systems 13~(2) (2020) 354--367.

\bibitem{rakotomamonjy2008bci}
A.~Rakotomamonjy, V.~Guigue, {BCI} {C}ompetition {III}: dataset {II}-ensemble
  of {SVM}s for {BCI} {P300} speller, IEEE Trans. Biomed. Eng. 55~(3) (2008)
  1147--1154.

\bibitem{zhong2020eeg}
P.~Zhong, D.~Wang, C.~Miao, {EEG}-based emotion recognition using regularized
  graph neural networks, IEEE Transactions on Affective Computing.

\bibitem{phan2018joint}
H.~Phan, F.~Andreotti, N.~Cooray, O.~Y. Ch{\'e}n, M.~De~Vos, Joint
  classification and prediction cnn framework for automatic sleep stage
  classification, IEEE Transactions on Biomedical Engineering 66~(5) (2018)
  1285--1296.

\bibitem{cecotti2010convolutional}
H.~Cecotti, A.~Graser, Convolutional neural networks for p300 detection with
  application to brain-computer interfaces, IEEE transactions on pattern
  analysis and machine intelligence 33~(3) (2010) 433--445.

\bibitem{li2021cross}
J.~Li, S.~Li, J.~Pan, F.~Wang, Cross-subject {EEG} emotion recognition with
  self-organized graph neural network, Frontiers in Neuroscience (2021) 689.

\bibitem{supratak2017deepsleepnet}
A.~Supratak, H.~Dong, C.~Wu, Y.~Guo, Deepsleepnet: A model for automatic sleep
  stage scoring based on raw single-channel {EEG}, IEEE Transactions on Neural
  Systems and Rehabilitation Engineering 25~(11) (2017) 1998--2008.

\bibitem{DBLP2015}
T.~Domhan, J.~T. Springenberg, F.~Hutter, Speeding up automatic hyperparameter
  optimization of deep neural networks by extrapolation of learning curves, in:
  IJCAI, 2015, pp. 3460--3468.

\end{thebibliography}

\end{document}